\documentclass[10pt]{article}
\usepackage{amsmath}
\usepackage{amsfonts}
\usepackage{amssymb}
\usepackage{xcolor}
\usepackage{color}
\usepackage[english]{babel}
\usepackage{graphics}
\usepackage{graphicx}
\begin{document}
\title{Decay Law of Relativistic Particles: \\
Quantum Theory Meets Special Relativity}
\author{K. Urbanowski\footnote{e--mail: K.Urbanowski@if.uz.zgora.pl}\\
Institute of Physics,
University of Zielona G\'{o}ra,  \\
ul. Prof. Z. Szafrana 4a,
65-516 Zielona G\'{o}ra,
Poland}
\maketitle
\begin{abstract}
Late time properties of moving relativistic particles are studied.
Within the proper relativistic treatment of the problem
we find decay curves of such particles and we show that late
time deviations of the survival probability of these particles
from the exponential form of the decay law, that is  the
transition times region between exponential
and non-expo\-nen\-tial form of the survival
amplitude, occur much earlier than it follows from the classical standard
approach boiled down to replace time $t$ by $t/\gamma_{L}$
(where $\gamma_{L}$ is the relativistic Lorentz factor) in the formula
for the survival probability. The consequence is that fluctuations of
the corresponding decay curves can appear much earlier
and much more unstable particles have a chance to survive
up to these times or later. It is also shown that fluctuations of
the instantaneous energy of the moving unstable particles has
a similar form as the fluctuations in the particle rest frame
but they are seen by the observer in his rest system much earlier than
one could expect replacing $t$ by $t/\gamma_{L}$ in the corresponding
expressions for this energy and that the amplitude of these fluctuations
can be even larger than it follows from the standard approach. All
these effects seems to be important when interpreting some accelerator
experiments with high energy unstable particles and the like
(possible connections of these effects with GSI anomaly are analyzed)
and some results of astrophysical observations.
\end{abstract}

\section{Introduction}

The problem of properties of unstable particles (states), their time
evolution and properties of the decay law has still not been definitely
solved as well as within quantum mechanics as within the quantum
field theory. There were published  plenty of papers in which various
aspects of this problem were analyzed and discussed. Particular attention
was focused on early and late time  properties of quantum decay processes.
It was shown that at these time regions classical exponential decay law
is unable to describe correctly a behavior of unstable quantum systems.
Early time deviations of the survival probability  from the exponential
form  lead to the so called Quantum Zeno and Anti--Zeno Effects
\cite{misra,chiu1,anti-zeno}. A conclusion that late time deviations
from the classical decay law have to take place in the case of quantum
decays follows from basic principles of the quantum theory: From the
postulate that spectrum of the total self--adjoint  hamiltonian $H$ of
the system containing unstable states has to be bounded from below
\cite{khalfin-2,fonda} it follows that at suitable late times the
quantum decay process must run more slowly than
any classical decay process described by an exponentially  decreasing
function, that is that the survival probability tends to zero as time $t$
goes to infinity more slowly than any exponentially decreasing function of time \cite{khalfin-2}.
There were many unsuccessful  attempts to  detect experimentally these
predicted  late time deviations (see, e.g., \cite{Wessner,Norman1}). Nevertheless
theoretical studies of this problem were still continued (see, e.g.
\cite{seke,parrot,lawrence,joichi,Nowakowski,Nowakowski2,jiitoh}).
Conclusions following from these studies were  applied successfully by Rothe and his group in
the experiment described  in \cite{rothe},
where the experimental evidence of deviations from the
exponential decay law at long times  was reported.
This result gave rise to another problem:
If (and how) deviations from the
exponential decay law at long times affect
the energy of the unstable state at this time region.
Analyzing the
transition times region between exponential
and non-expo\-nen\-tial form of the survival
amplitude it  has been shown in \cite{plb-2014}  that the instantaneous
energy of the unstable particle can take very
large values, much larger than the energy of
this state for times from the exponential
time region.  It has been shown that this
purely quantum  mechanical effect may
force relativistic unstable particles
to emit electromagnetic--, $X$-- or
$\gamma$--rays  at some time intervals from  the
transition time regions.
It has been  hypothesized in \cite{plb-2014} that this effect may
be responsible for some astrophysical effects such as
cosmic  radio,  $X$-- or $\gamma$--rays bursts, etc.

The problem is that from the point of view of a frame of reference  in which
the time evolution of the unstable system takes place
the Rothe experiment as well as the properties of unstable states
discussed usually in the  literature and mentioned above
refer to the rest coordinate system of the
unstable system considered. Astrophysical sources of unstable particles emit
them  with relativistic or ultra--relativistic velocities in relation to an external
observer. The question is what effects can be observed
by an external observer when the unstable
particle, say $\phi$,   which survived up to the transition
times region or longer is moving with a relativistic velocity in
relation to this observer. The related question is how the time dilation formula
being the classical physics formula works in the case of quantum decay
processes, and especially how it works at late times when the main contribution
to the survival probability comes from the  nonexponential corrections,
which are purely quantum nature
and are absent in decay laws considered in classical physics.
Such and similar problems seems to be extremely important because  quantum
decay process of moving relativistic particles are the place where quantum
theory meets special relativity, which is the classical theory.
Many authors tried to show that time dilation formula works under some approximations
simplifying general analysis of properties of survival amplitudes under
Lorentz transformations:
Unfortunately these simplifications seem to be valid only for relative
small times (see e.g., \cite{exner}). More general analysis
based on the correct use of the dependence of the energy  of the moving relativistic
unstable particle on its rest mass and its momentum leads to conclusions
that the classical decay law taking into account time dilation may differ from
the correct quantum decay law calculated for moving relativistic particles
at late times (see \cite{khalfin-1,stefanovich,shirkov}) and that this
difference  growths at very late times as $t$ increases \cite{shirkov}.
Taking into account experiments realized in the Earth laboratories these
and similar problems may seem to be
of a very little importance  and purely academic. Nevertheless it seems that
the proper interpretation of
all results of the accelerator experiments with unstable objects of extremely
large energies is impossible without knowledge of  properties of survival
probabilities at all times, including transition and very late times, and when
the transition times begin. On the other hand the correct interpretation of
observational results and possible effects caused by unstable particles
emitted by astrophysical sources is rather  impossible without detailed
knowledge of the late time
behavior of these moving particles.
It is because astrophysical sources  produce such huge number of particles that
many of them are able to survive up to transition times or even much
longer (see \cite{plb-2014} and references therein).

The letter is organized as follows: A general late time properties of moving,
relativistic unstable particles are analyzed in Sec. 2. Results of numerical
calculations for a given model  are presented in the graphical form in Sec. 3.
Sec. 4 contains a discussion and final remarks.

\section{Late time properties of moving unstable particles}

Let us analyze the problem of  determining the decay law, i.e.,
the non-decay probability (or the survival probability) ${\cal P}(t)$ of
the moving unstable particle with nonzero momentum $p = |\vec{p}| \neq 0$.
From the standard, text book  considerations one finds that if
the decay law of the unstable particle in rest
($p = 0$) has the exponential form ${\cal P}_{0}(t) = \exp\,[- {\it\Gamma}_{0}\,t]$
then the decay law of the moving particle with momentum $p \neq0$ is
${\cal P}_{p}(t) = \exp\,[-\,{\it\Gamma}_{0}\,\frac{m_{0}}{\sqrt{p^{2} + m_{0}^{2}}}\,t]\,
\equiv \,\exp\,[-\,{\it\Gamma}_{0}\,\frac{t}{\gamma_{L}}]$, where
$m_{0}$ is the rest mass of the particle and $\gamma_{L}$ is the
relativistic Lorentz factor, $\gamma_{L} \equiv \frac{1}{\sqrt{1 - \beta^{2}}},
\;\;\beta = v/c$, $v$ is the velocity of the particle.  It is almost
common belief that this equality is valid for any $t$.
Similar belief concerns a more general relation between probability amplitudes
\begin{equation}
|a_{p}(t)|^{2} = |a_{0}(t/\gamma_{L})|^{2}, \label{a-p=a-0}
\end{equation}
where $a_{0}(t)$ is the  probability amplitude of finding the system at the
time $t$ in the initial state $|\phi\rangle$ prepared at time $t_{0}
= 0$ and it refers to the particle rest coordinate system,
$a_{0}(t) = \langle \phi|\phi (t) \rangle$ and $|\phi (t) \rangle = \exp\,[-itH]\,|\phi\rangle$,
$H$ is the selfadjoint Hamiltonian of the total system under considerations,
and $a_{p}(t) = \langle p; \phi|\phi (t);p \rangle$, where $|\phi (t);p \rangle =
\exp\,[-itH]\,|\phi;p\rangle$ and $|\phi;p\rangle$ is the state vector of the
moving unstable particle $\phi$  and having a momentum $p \neq 0$
and it is obtained by expanding $|\phi\rangle$
in the basis of common eigenvectors
of $H$ and of the momentum operator
${\bf P}$  (for details see \cite{shirkov}).
 The corresponding
survival probabilities are defined as follows:
${\cal P}_{0}(t) = |a_{0}(t)|^{2}$, ${\cal P}_{p}(t) = |a_{p}(t)|^{2}$.
Equation (\ref{a-p=a-0}) represents the so called Einstein time dilation.
Some, approximate model calculations
show that time dilation in the form expressed by Eq. (\ref{a-p=a-0})
does not hold exactly in the case of moving unstable particles. Although
in \cite{exner} it was found within the quantum field theory considerations
that $a_{p}(t) = a_{0}(t/\gamma_{L})$ but this relation
was obtained there as the approximate one and valid only for
not a very large number of lifetimes.
Similar reservations
in relation to
the property (\ref{a-p=a-0}) can be found in \cite{khalfin-1,stefanovich,shirkov}.
For the more detailed analysis of the problem
we need the exact form of the amplitudes $a_{0}(t)$ and $a_{p}(t)$ for
all $t$ (and thus corresponding survival probabilities), if not in the general case,
then  at least for a reasonable realistic model of the moving unstable particle.

From basic principles of quantum theory it is known that the
amplitude $a_{0}(t)$, and thus the decay law ${\cal P}_{0}(t)$ of the
unstable state $|\phi\rangle$, are completely determined by the
density of the energy distribution $\omega({\cal E})$ for the system
in this state \cite{Fock,khalfin-2}, or, equivalently by the density of
the mass distribution $\omega (m)$ \cite{khalfin-1,shirkov}.
There is (in $\hbar = c = 1$ units)
\begin{equation}
H|m;0\rangle = m|m;0\rangle,\;\;\;m\in \sigma_{c}(H), \label{m;0}
\end{equation}
(where $|m;0\rangle = |m;p=0\rangle$) in the rest coordinate system, and
\begin{equation}
{\bf P}|m;0\rangle = 0, \;{\rm and}\;\;\; {\bf P}|m;p\rangle =
\vec{p}|m;p\rangle, \;({\rm for}\; \vec{p}\neq 0). \label{P;0}
\end{equation}
Thus
\begin{equation}
|\phi\rangle = \int_{\mu_{0}}^{\infty}\,c(m) |m;0\rangle\, dm, \label{phi-spec}
\end{equation}
and
\begin{equation}
a_{0}(t) \stackrel{\rm def}{=} \langle \phi| e^{\textstyle{-itH}}|\phi\rangle
= \int_{\mu_{0}}^{\infty} |c(m)|^{2}\;
e^{\textstyle{-\,i\,m\,t}}\,d{m},
\label{a-spec}
\end{equation}
where  $\mu_{0}$ is the lower bound of the spectrum $\sigma_{c}(H)$ of $H$.
The density of the mass distribution is defined as
follows $\omega(m) \stackrel{\rm def}{=}|c(m)|^{2} > 0$.
 A reasonable simplified representation of the density of the mass
distribution is to choose the Breit--Wigner form for $\omega (m)$, which under
rather general condition approximates sufficiently well real
systems \cite{stefanovich,fonda,nowakowski3},
\begin{equation}
\omega (m) = \omega_{BW}(m) \stackrel{\rm def}{=} \frac{N}{2\pi} \it\Theta (m - \mu_{0})
\frac{{\it\Gamma}_{0}}{(m - m_{0})^{2} +
(\frac{\it\Gamma_{0}}{2})^{2}}, \label{omega-BW}
\end{equation}
where $N$ is a normalization constant and ${\it\Theta} (m)$ is the
unit step function. Inserting this $\omega (m)$ into (\ref{a-spec})
one finds that for very late times the amplitude $a_{0}(t)$ has
the following form (see, eg. \cite{urbanowski-2008})
\begin{eqnarray}
{a_{0}(t)\vline}_{\, t \rightarrow \infty} &\simeq & N e^{\textstyle -
i\,h_{0}\,t}\,+\nonumber \\&& - i \frac{N}{2
\pi}\; e^{\textstyle{-it\mu_{0}}}\;\frac{{\it\Gamma}_{0}}{|\,h_{0}\,-\,\mu_{0}\,|^{\,2}}
 \, \frac{1}{t}
\,+ \ldots\;\;, \label{a(t)-as}\\
&=& a_{0}^{exp}(t) + a_{0}^{lt}(t), \label{a-exp+a-lt}
\end{eqnarray}
where $h_{0} \equiv m_{0} - \frac{i}{2}\, {\it\Gamma}_{0}$ and
$a_{0}^{exp}(t) =  N\,\exp\,[-i\,h_{0}\,t] $, $a_{0}^{lt}= a_{0}(t) - a_{0}^{ex}(t)$.
The transition time region denotes times $t$ when the contributions
of $a_{0}^{exp}(t)$ and $a_{0}^{lt}(t)$ into the survival probability
${\cal P}_{0}(t)$ begin to be of the comparable order. To this time
region belongs times $t \sim T$, where $T$ is a solution of the following equation
\begin{equation}
|a_{0}^{exp}(t)|^{2} = |a_{0}^{lt}(t)|^{2}, \label{a-exp=a-lt}
\end{equation}
which in the considered case reads,
\begin{equation}
e^{\textstyle - {\it\Gamma}_{0}\,t}\; =\;
\frac{1}{4
\pi^{2}}\;\frac{({\it\Gamma}_{0})^{2}}{|h_{0}\,-\,\mu_{0}\,|^{\,4}} \;
\frac{1}{t^{2}}, \label{T-0}
\end{equation}
or, equivalently,
\begin{equation}
{\it\Gamma}_{0}\,t \;=\;\ln\Big[(2\pi)^{2}\,
\frac{|h_{0}-\mu_{0}|^{4}}{{\it\Gamma}_{0}^{4}}\,\Big]\;+\;2\,\ln[{\it\Gamma}_{0}\,t].
\label{T-0-1}
\end{equation}
The very approximate asymptotic solution,
$T_{0}$, of this equation for $\frac{m_{0}}{{\it\Gamma}_{0}}\,\gg\, 1$ (in general for
$\frac{m_{0}}{{\it\Gamma}_{0}}\,\rightarrow \,\infty$)
has the form
\begin{equation}
\frac{T_{0}}{\tau_{0}} \gtrsim  2\,\ln\,(4\pi) \,+\, 4\,
\ln\Big( \frac{m_{0}-\mu_{0}}{{\it\Gamma}_{0}} \Big)
\,+\,\ldots\;\; ,
\label{T-0-2}
\end{equation}
where  $\tau_{0} \equiv 1/{\it\Gamma}_{0}$ is a mean
lifetime of the unstable particle considered in its rest frame.
Results (\ref{a(t)-as}) and (\ref{T-0-1}), (\ref{T-0-2})
follow from (\ref{a-spec}) and (\ref{omega-BW}) and
refer to the rest coordinate system. Now we should compare them with
analogous results
obtained for  the moving unstable particle.

So, let us find  the probability amplitude, $a_{p}(t)$, of the moving unstable
particle relative to rest coordinate system of the observer $\cal O$
and having constant momentum $\vec{p}$ (here $p = |\vec{p}|$) measured by  $\cal O$.
If $\Lambda$ denotes the Lorentz transformation from the reference frame,
where the momentum of unstable particle considered is zero, $\vec{p} =0$,
into the frame where  the momentum of this particle is non--zero,
$\vec{p} \neq 0$ or, equivalently,  where its velocity $\vec{v}$
equals $\vec{v} =\vec{v}_{p} \equiv \frac{\vec{p}}{m\,\gamma_{L}} \neq 0$,
(the momentum $\vec{p}$ is given), then
\begin{equation}
|m;p\rangle = U(\Lambda) |m;0\rangle,
\end{equation}
(where $ U(\Lambda)$ is a unitary representation  of the transformation
$\Lambda$  and acts in the Hilbert space of states
$|\phi\rangle = |\phi;p=0\rangle, |\phi;p\rangle$), and
\begin{eqnarray}
|\phi;p\rangle = U(\Lambda) |\phi\rangle &=&
\int_{\mu_{0}}^{\infty}c(m)\,U(\Lambda) |m;0\rangle\,dm \nonumber \\
&\equiv&
\int_{\mu_{0}}^{\infty}c(m)\, |m;p\rangle\,dm. \label{m;p}
\end{eqnarray}
Operators $H, \bf{P}$ form a 4--vector $P_{\nu}=(P_{0},{\bf P})
\equiv  (H,{\bf P})$. Therefore
$U^{+}(\Lambda) P_{\nu}U(\Lambda) = \Lambda_{\nu \lambda}\,P_{\lambda}$, where  $\lambda, \nu =0,1,2,3$,
(see, e.g., \cite{gibson}, Chap. 4) and thus \cite{gibson}
\begin{equation}
U^{+}(\Lambda) P_{0} U(\Lambda) = \gamma_{L}(P_{0} + \vec{v}_{p}\cdot{\bf P}). \label{UPU}
\end{equation}
From this last relation it follows that vectors $|m;p\rangle$
are also eigenvectors for the Hamiltonian $ H \equiv P_{0}$.
Indeed using (\ref{m;0}), (\ref{m;p}) and (\ref{UPU}) one finds that
\begin{equation}
H|m;p\rangle = m\,\gamma_{L}\;|m;p\rangle. \label{H-m;p}
\end{equation}
Now keeping in mind that the momentum $\vec{p}$ is given and constant,
which means that in this case the product $m\,\gamma_{L}$ can be expressed as follows
$m\,\gamma_{L}\,\equiv\,\sqrt{p^{2} + m^{2}}$, one concludes that simply
\begin{equation}
H|m;p\rangle \,= \,\sqrt{p^{2} + m^{2}}\,|m;p\rangle. \label{H-m;p-0}
\end{equation}
So we  see finally that in the considered case of the moving unstable
particle with a constant momentum $\vec{p} \neq 0$
we obtain the following formula for the probability amplitude $a_{p}(t)$,
\begin{equation}
a_{p}(t) \equiv \langle p;\phi|e^{\textstyle{-itH}}|\phi;p\rangle \,
= \,\int_{\mu_{0}}^{\infty} \omega(m)\;
e^{\textstyle{-\,i\sqrt{p^{2}+m^{2}}\;\,t}}\,d{m}, \label{a-p-spec}
\end{equation}
instead of  the expression (\ref{a-spec}) for the probability amplitude $a_{0}(t)$
with the same $\omega (m)$. (For more details, a discussion and explanations see,
e.g. \cite{khalfin-1,stefanovich,shirkov}). This representation
of the amplitude $a_{p}(t)$ is valid for any $p$ and for
$p \rightarrow 0$ it transforms into (\ref{a-spec}).

Inserting (\ref{omega-BW}) into (\ref{a-p-spec})
and then assuming for simplicity that $\mu_{0} =0$
enables one to reproduce calculations performed by Shirkov in \cite{shirkov}
and  to obtain asymptotic form of $a_{p}(t)$, which within
the use units $\hbar \neq 1 \neq c$ reads as follows
\begin{eqnarray}
a_{p}(t) &\simeq & Ne^{\textstyle{-\frac{i}{\hbar}\Big[(1-\alpha)\sqrt{(cp)^{2}
+ (m_{0}c^{2})^{2}}-\frac{i}{2}\frac{(1+\alpha)}{\gamma_{L}}{\it\Gamma}_{0} \Big]t}} \nonumber \\
&& -\frac{N}{2\sqrt{2\pi}}\frac{{\it\Gamma}_{0}}{m_{0}c^{2}}\frac{cp}{m_{0}c^{2}}\frac{1}{\sqrt{cp\frac{t}{\hbar}}}
e^{\textstyle{-\frac{i}{\hbar}cpt}}e^{\textstyle{i\frac{\pi}{4}}}+\ldots, \label{a-p-as}\\
& \stackrel{\rm def}{=}& a_{p}^{exp}(t) + a_{p}^{lt}(t), \label{a-p-exp+a-p-lt}
\end{eqnarray}
where
\begin{eqnarray}
\alpha & = & \frac{1}{8}\frac{{\it\Gamma}_{0}^{2}}{(cp)^{2}+(m_{0}c^{2})^{2}}\,\frac{(cp)^{2}}{(cp)^{2}+(m_{0}c^{2})^{2}} \nonumber\\
&\equiv& \frac{1}{8}\Big(\frac{{\it\Gamma}_{0}}{m_{0}c^{2}}\Big)^{2}\;\frac{\gamma_{L}^{2} - 1}{\gamma_{L}^{4}}, \label{alpha}
\end{eqnarray}
(for details see \cite{shirkov}: A substitution of $\hbar = 1 = c$ into (\ref{a-p-as}),
(\ref{alpha})  yields formulae obtained there).  Probability amplitudes $a_{p}^{exp}(t), \;
a_{p}^{lt}(t)$ denote the exponential and the late time nonexponential parts of the amplitude $a_{p}(t)$.
The relation (\ref{a-p-as}) is valid if $\frac{m_{0}c^{2}\,t}{\hbar}\,
\gg\,1$ and $\frac{cp\,t}{\hbar}\,\gg\,1$ and therefore the limit
$p \rightarrow 0$ can not be performed   in (\ref{a-p-as}). From (\ref{alpha}) it
follows that $\alpha$ reaches its maximal value for $\gamma_{L} = \sqrt{2}$.

Using the equation $|a_{p}^{exp}(t)|^{2} = |a_{p}^{lt}(t)|^{2}$ one can
find the time $T_{p}$
defining the transition times
region for moving unstable particles. The explicit form of this equation  looks as follows
\begin{equation}
e^{\textstyle{-\,\frac{1+\alpha}{\gamma_{L}}\,\frac{{\it\Gamma}_{0}t}{\hbar}}}
= \frac{1}{8\pi}\,\Big(\frac{{\it\Gamma}_{0}}{m_{0}c^{2}}\Big)^{2}
\,\frac{cp}{(m_{0}c^{2})^{2}}\,\frac{\hbar}{t}. \label{a-p-exp=a-p-lt-a}
\end{equation}
A very approximate asymptotical solution, $T_{p}$, of
Eq. (\ref{a-p-exp=a-p-lt-a}) has the following form
\begin{eqnarray}
\frac{1 + \alpha}{\gamma_{L}}\,\frac{T_{p}}{\tau_{0}} & \gtrsim &
\ln\,\Big[8 \pi \Big(\frac{m_{0}\,c^{2}}{{\it\Gamma}_{0}}\Big)^{4}\,\frac{{\it\Gamma}_{0}}{cp}
\,\frac{\gamma_{L}}{1+\alpha}\Big]\,+\,\ldots \nonumber\\
&\equiv & \ln\,(8\pi)\;+\;3\,\ln\,\Big(\frac{m_{0}c^{2}}{{\it\Gamma}_{0}}\Big) \nonumber \\
&& -\;\ln\,[(1+\alpha)\beta]\;+\;\ldots \label{T-p}
\end{eqnarray}
The limit $p\rightarrow 0$ (or, equivalently, $\gamma_{L} \rightarrow 1$
or $\beta \rightarrow 0$) is not applicable to the relation (\ref{T-p}).
It is  because  (\ref{T-p}) is a solution of Eq. (\ref{a-p-exp=a-p-lt-a})
following from the relation (\ref{a-p-as}) which holds  under the condition
that limitations formulated after formulae (\ref{a-p-exp+a-p-lt}), (\ref{alpha}) take place.

As it was mentioned earlier it is common belief that in order to obtain
the survival probability, ${\cal P}_{p}(t)$ of the moving relativistic unstable
particle it is sufficient to replace time $t$ in the survival probability
${\cal P}_{0}(t) = |a_{0}(t)|^{2}$ of decaying particle in  its rest
coordinate system by $t'=t/\gamma_{L}$. Such a  "recipe" leads to the
conclusion that in the case of moving unstable particles the transition
time $T'$ corresponding with the solution $T_{0}$ of Eqs. (\ref{a-exp=a-lt}), (\ref{T-0})
can be found replacing $t$ in Eq. (\ref{T-0}) by $t'=t/\gamma_{L}$. The solution,
$T'/\tau_{0}$, of such a problem  has an analogous form as the solution,
$T_{0}/\tau_{0}$, (\ref{T-0-2}) with $T'=T_{0}/\gamma_{L}$ replacing $T_{0}$
in (\ref{T-0-2}). Such obtained formula for the transition time $T'$ of
a moving decaying particle differs significantly
from  the solution $T_{p}$, (\ref{T-p}),
of Eq. (\ref{a-p-exp=a-p-lt-a}), which was obtained using proper relativistic expression
(\ref{a-p-spec}) for the probability
amplitude $a_{p}(t)$:  If to assume that $\mu_{0} = 0$ in (\ref{T-0}), (\ref{T-0-2})
and if the Lorentz factor $\gamma_{L}$ is suitable large, $\gamma_{L} \gg 1$,
then within the considered model to a very good approximation,
\begin{eqnarray}
\frac{T'}{\tau_{0}}\,-\,\frac{T_{p}}{\gamma_{L}\;\tau_{0}} & \equiv &
\frac{T_{0}}{\gamma_{L}\;\tau_{0}}\,-\,\frac{T_{p}}{\gamma_{L}\;\tau_{0}} \nonumber\\
& \gtrsim & \ln \, (2\pi) \; + \; \ln \, \big( \frac{m_{0}}{{\it\Gamma}_{0}} \big).
\label{T'-T-p}
\end{eqnarray}

This result means among others that  fluctuations of the instantaneous
energy, ${\cal E}_{\phi}(t)$, of a moving unstable particle $\phi$
mentioned in Sec. 1 and
discussed in \cite{plb-2014} begins much earlier than it could be
expected assuming that the relation (\ref{a-p=a-0}) holds  for
relativistic unstable particles at all times. The instantaneous energy,
${\cal E}_{\phi}^{p}(t)$, of the moving particle with momentum $p$ is defined
analogously as the instantaneous energy in the particle rest
system (see \cite{urbanowski-2008,urbanowski-1-2009,pra}):
${\cal E}_{\phi}^{p}(t) \,=\,\Re\,[h^{p}(t)]$, where
\begin{equation}
h^{p}(t) = i\,\frac{\partial a_{p}(t)}{a_{p}(t)}, \label{h-p}
\end{equation}
is the effective hamiltonian governing the time evolution of the
particle considered. In the general case assuming the form of the
density $\omega (m)$ and starting from the relation (\ref{a-p-spec})
real and imaginary parts of $h^{p}(t) $ can be found numerically.
For the model considered the asymptotic late time form, $h^{p}_{as}(t)$,
of $h^{p}(t)$ can be easily found  using $a_{p}^{lt}$ given by formulae
(\ref{a-p-as}), (\ref{a-p-exp+a-p-lt}). There is
\begin{equation}
h^{p}_{as}(t) \simeq i\,\frac{\partial a_{p}^{lt}(t)}{a_{p}^{lt}(t)} =
p\; - \;\frac{i}{2}\,\frac{1}{t}\,+\,\ldots\;\;, \;\;(t \rightarrow \infty).
\label{h-as}
\end{equation}
This means that within the model considered
\begin{equation}
{\cal E}^{p}_{as}\; \simeq \; p\;+\;\ldots , \;\;{\rm and,}\;\;{\it\Gamma}^{p}(t) = \frac{1}{t}\,+ \,\ldots ,\
(t \rightarrow \infty),
\label{E-as}
\end{equation}
where ${\it\Gamma}^{p}(t) = - 2\,\Im\,[h^{p}(t)] \equiv \,-
\,\frac{1}{{\cal P}_{p}(t)}\,
\frac{\partial {\cal P}_{p}}{\partial t}$ is the instantaneous decay rate
(or,  using units $\hbar \neq 1 \neq c$,  $h_{p}^{as}(t)
\simeq \,cp\, - \,\frac{i}{2}\,\frac{\hbar}{t}\,+\,\ldots$ and
 ${\cal E}^{p}_{as}\, \simeq\, cp\,+\,\ldots$,
 $\;\;{\it\Gamma}^{p}(t) =\,\hbar/t\,+\ldots$).

\section{Numerical results}

Asymptotic late time forms $a_{0}^{lt}, a_{p}^{lt}$ of
the probability amplitudes $a_{0}(t)$, $a_{p}(t)$ and thus
corresponding survival probabilities ${\cal P}_{0}(t), {\cal P}_{p}(t)$
and instantaneous energies ${\cal E}^{0}_{as}(t) = \Re\,[h^{0}_{as}(t)]$,
(where,  $h^{0}(t) = i [ \partial a_{0}(t)/\partial t] (a_{0}(t))^{-1}$
and ${\cal E}^{0}(t) = \Re\,[h^{0}(t)]$),
 and  ${\cal E}^{p}_{as}(t)$ are relatively
easy to find analytically for times $t \gg T_{0}$ and $t \gg T_{p}$ even in
the general case as it was shown in
\cite{urbanowski-1-2009}.
It is rather impossible to find a transparent and readable form of these quantities
at time regions, when  $t \sim T_{0}$ or $t \sim T_{p}$.
For the model considered (\ref{omega-BW}) it can be done numerically.
The results presented in this Section have been obtained
assuming for simplicity that the minimal mass (energy)
$\mu_{0}$ appearing in the formula (\ref{omega-BW}),  and thus
also in (\ref{a-spec}) and (\ref{a-p-spec}), is equal to zero,
$\mu_{0} = 0$ (or $E_{min} = \mu_{0}\,c^{2} =0$). Calculations
have been performed  for some chosen $\frac{m_{0}}{{\it\Gamma}_{0}}$
and $\frac{p}{{\it\Gamma}_{0}}$.
Performing calculations  particular attention was paid to
the form of the survival probability, i. e. of the decay
curve, and of the instantaneous energy ${\cal E}_{\phi}(t)$
for times $t$ belonging to the most interesting time regions:
For transition times $t \sim T_{0}$ and $t \sim T_{p}$
and for times $t \gg T_{0}$ and $t\gg T_{p}$ when the late
time asymptotic parts of the probability amplitudes
are dominant. Results are presented graphically
below in Figs \ref{fi1}, \ref{fi2}.
\begin{figure}[h!]
\begin{center}
\includegraphics[width=68mm]{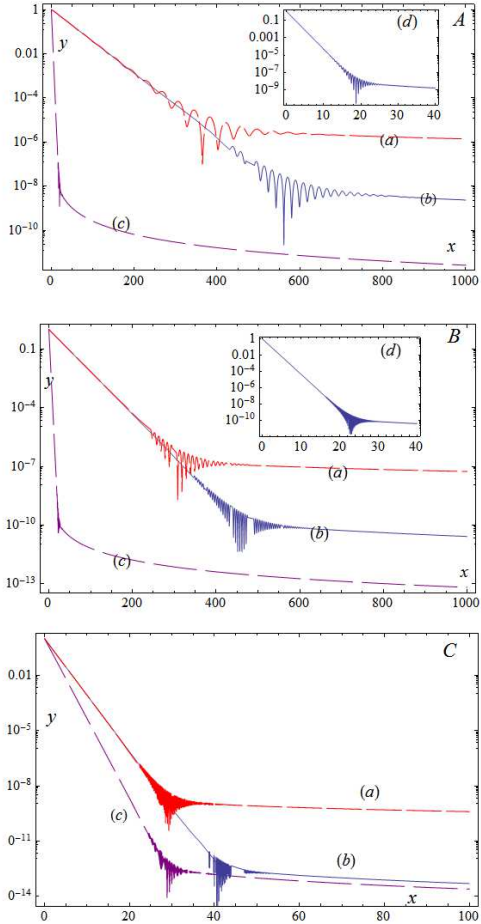}
\caption{Axes:
$x =t / \tau_{0}$ --- time is measured in lifetimes $\tau_{0}$;  $y$ --- survival probabilities (the logarithmic scale).  In all panels: $(a)$ the survival probability ${\cal P}_{p}(t)$, $(b)$ the survival probability
${\cal P}_{0}(t/\gamma_{L})$, $(c)$ the survival probability ${\cal P}_{0}(t)$ and $(d)$ is the enlarged part of $(c)$ showing  the form of the decay curve ${\cal P}_{0}(t)$ for times $t\sim T_{0}$ belonging to the transition times region. Panel $A$:  $\gamma_{L} \simeq 30.0167$ which corresponds to $m_{0}/{\it\Gamma}_{0} = 10$ and $p/{\it\Gamma}_{0} = 300$; panel $B$: $\gamma_{L} \simeq 20.025$, ($m_{0}/{\it\Gamma}_{0} = 25$ and $p/{\it\Gamma}_{0} = 500$); panel $C$:  $\gamma_{L} = \sqrt{2}$ which corresponds to $m_{0}/{\it\Gamma}_{0} = 100$ and $p/{\it\Gamma}_{0} = 100$.}
  \label{fi1}
\end{center}
\end{figure}

Results presented in these Figures enable us to
compare decay curves of a moving relativistic
unstable particle obtained within a correct relativistic treatment of
the evolving in time $t$  and moving
particle having
certain momentum $p$ seen by
an observer $\cal O$ in his rest system with those followed from
the standard classical reasoning that in order to obtain relativistic effects for
 such a particle
it is sufficient to replace $t$ by $t'=t/\gamma_{L}$ (see (\ref{a-p=a-0})).
Note that these results are in perfect agreement with analytical estimations
(\ref{a-p-as}), (\ref{T-p}) and (\ref{T'-T-p}) performed for the model considered.

A similar comparison can be done in the case of the instantaneous energy
${\cal E}^{p}(t)$ of the moving unstable particle, which was discussed in \cite{plb-2014}
where it was shown that fluctuations of this energy are responsible for
a possible emission of the electromagnetic radiation by moving charged unstable particles.
Changes of these energies relative to the energy of the moving particle at the
canonical decays time region (where the survival probability has the
exponential form), $E_{0}(p) = \sqrt{p^{2} + m_{0}^{2}}$, are presented
 in Fig \ref{fi2} in the form of ratios: $\kappa_{p}(t) \stackrel{\rm def}{=} \frac{{\cal E}^{p}(t) - {\cal E}^{p}_{as}}{E_{0}(p) - {\cal E}^{p}_{as}}$, (here ${\cal E}^{p}_{as} = p$ --- see (\ref{E-as})), and
$\kappa_{0}(t) = \frac{{\cal E}^{0}(t) - {\cal E}^{0}_{as}}{E_{0} - {\cal E}^{0}_{as}}$, (here: $E_{0} = {E_{0}(p)\,\vline}_{\,p=0} \equiv m_{0}$, ${\cal E}^{0}_{as}= \;E_{min} \equiv \mu_{0} =0$ and ${\cal E}^{0}(t) = \Re\,[h^{0}(t)]$, $h^{0}(t) = {h^{p}(t)\vline}_{\,p=0}$).

\begin{figure}[h!]
\begin{center}
\includegraphics[width=68mm]{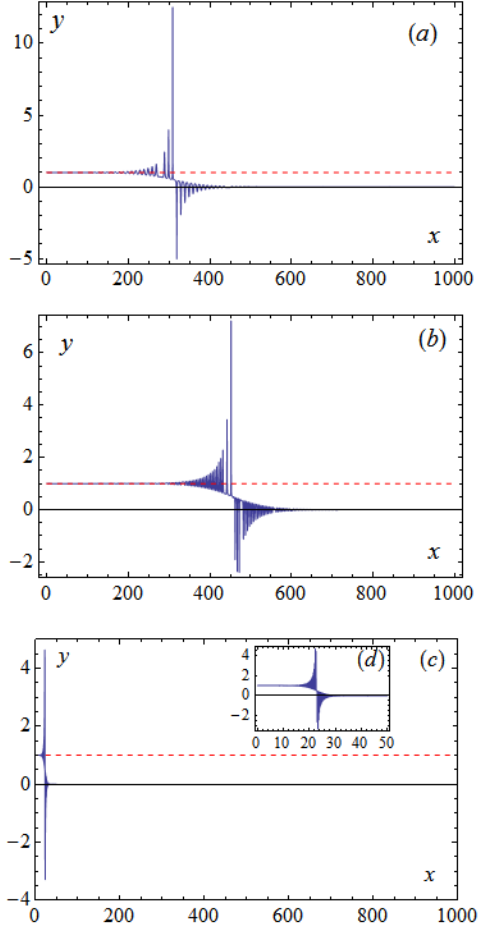}\\
\caption{
Axes: $x =t / \tau_{0} $, and  $y$ --- instantaneous energies:  panel $(a)$ $\kappa_{p}(t)$; panel $(b)$ $\kappa_{0}(t/\gamma_{L})$ and panel (c) $\kappa_{0}(t)$. The horizontal dashed line denotes the values of $\kappa$-s equal to $1$.
Here $\gamma_{L} = 20.025$ which corresponds to $m_{0}/{\it\Gamma}_{0} = 25$ and $p/{\it\Gamma}_{0} = 500$. }
  \label{fi2}
\end{center}
\end{figure}

\section{Discussion and final remarks}

Reasonable and physically acceptable models of unstable particles
defined by means of the density $\omega (m)$  usually have the following form:
$\omega (m) = \omega_{BW}(m)\;\times\;(m-\mu_{0})^{\lambda}\,\times\, f(m)$,
where $\lambda \geq 0$ and $f(m)$ is a form--factor ---  it is
a smooth function going to zero as $ m  \rightarrow \infty$ and
it has no threshold and no pole. It appears that
a behavior of the amplitudes,
$a_{0}(t)$, defined by such a density $\omega (m)$
and by $\omega_{BW}(m)$ as functions of time $t$
is very similar (see \cite{nowakowski3,fonda}). So conclusions
following from the results obtained in Sec. 3 and 4 seems to be  sufficiently general.

\begin{figure}[h!]
\begin{center}
\includegraphics[width=68mm]{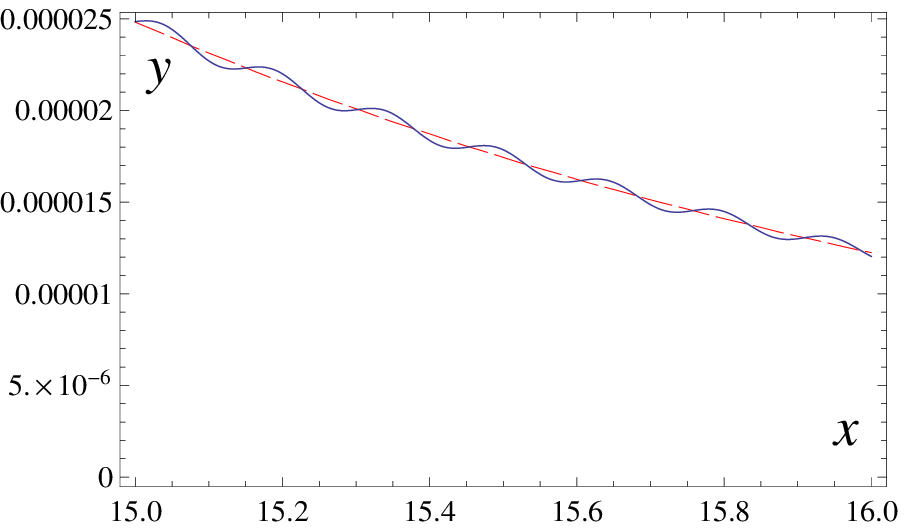}
\caption{  The case $m_{0}/{\it\Gamma}_{0} = 100$ and $p/{\it\Gamma}_{0} = 100$ (which gives $\gamma_{L} = \sqrt{2}$): The enlarged part of the decay curves $(a)$ and $(b)$ in Panel $(C)$ of  Fig. \ref{fi1}.
Axes:
$x =t / \tau_{0}$, and  $y$ --- survival probabilities: Solid line --- the survival probability ${\cal P}_{p}(t)$; Dashed line --- the survival probability
${\cal P}_{0}(t/\gamma_{L})$.}
  \label{fi3}
\end{center}
\end{figure}

Results presented in Sec. 3 and 4 show that the relation (\ref{a-p=a-0}) can be
considered as a sufficiently accurate only for no more than a few lifetimes
$\tau_{0}$ and that   the supposition  that (\ref{a-p=a-0}) holds  for
times $t \gg \tau_{0}$ is wrong. It is because the assumption
that ${\cal P}_{p} (t) \stackrel{\bf ?}{=} {\cal P}_{0}(t/\gamma_{L})$ is
the classical physics relation.
An extension of it to quantum decay processes does not lead to
a significant error only for times $t$ when classical and quantum decay laws
have a similar classical form, that is the exponential form. When quantum effects
force the survival probability ${\cal P}_{p}(t)$ to behave nonclassically then the relation
(\ref{a-p=a-0}) is wrong and it may lead to the incorrect interpretation
of decays of relativistic particles. Such a possible  hypothetical situation is presented
in Fig. \ref{fi3}: The temporal behavior of the real decay process of a relativistic
particle at  time intervals containing times $t$ significantly smaller than
$T_{p}$ is described  by the survival probability ${\cal P}_{p}(t)$ and it is shown
by the solid line in this Figure, whereas the dashed line represents
${\cal P}_{0}(t/\gamma_{L})$ and according to (\ref{a-p=a-0}) it is usually
interpreted as the correct illustration of the decay process of such a particle.

In Fig. \ref{fi1} are compared decay curves, that is survival probabilities
${\cal P}_{p}(t)$, obtained within the correct relativistic treatment of
 evolving in time and moving unstable particles with a given momentum $p$
relative to the rest system of the observer  $\cal O$ as seen by this observer, with those obtained assuming
the validity of the standard classical reasoning that in order to get
decay curves of such particles it is sufficient to replace time $t$ in ${\cal P}_{0}(t)$
obtained using (\ref{a-spec}) by $t'= t/\gamma_{L}$, that is it is enough to
consider ${\cal P}_{0}(t/\gamma_{L})$ instead of ${\cal P}_{0}(t)$. Numerical
results presented in this Figure are entirely consistent
with the analytical results obtained in Sec. 2.

From (\ref{T-0-2}), (\ref{T-p}) and (\ref{T'-T-p}), or comparing decay curves $(a)$ and $(b)$
in Fig. \ref{fi1}, one can conclude that
in the case of moving particles the transition time regions
begin much more earlier than one could expect using the relation (\ref{a-p=a-0}).
For some combinations of $m_{0}/{\it\Gamma}_{0}$ and $p/{\it\Gamma}_{0}$
the transition times regions can begin even earlier  in the case of moving particles,
$p\neq 0$, than such a time region in the case of the particle observed
in its rest system, $p=0$, (see Fig. \ref{fi1}, Panel $C$.
A consequence of this fact
is that ${\cal P}_{p}(T_{p})
\gg {\cal P}_{0}(T_{0}/\gamma_{L})$ for times $t \geq T_{p}$. What is more,
from  results obtained in Sec. 2 and 3 it is seen that correctly obtained
survival probability ${\cal P}_{p}(t)$ tends to zero as $t \rightarrow \infty$
much more slowly than ${\cal P}_{0}(t/\gamma_{L})$: Within the model
considered ${\cal P}_{p}(t) \sim 1/t $ and  ${\cal P}_{0}(t/\gamma_{L})
\sim 1/t^{2}$ for $t\rightarrow \infty$ which confirms the conclusions presented in \cite{shirkov}.
So if the initial number of unstable particles was ${\cal N}_{0}$,
then  the real number of moving unstable particles ${\cal N}(T_{p})$ which had a chance
to survive up to time $t \sim T_{p}$ or later  and which were registered by
the observer $\cal O$ is much greater
than the corresponding number
${\cal N}(T_{0}/\gamma_{L})$ obtained assuming the validity of (\ref{a-p=a-0}):
There is ${\cal N}(T_{p}) = {\cal P}_{p}(T_{p})\,{\cal N}_{0}\,\gg \,
{\cal N}(T_{0}/\gamma_{L}) = {\cal P}_{0}(T_{0}/\gamma_{L})\,{\cal N}_{0}$.
 A similar conclusion holds for times $t > T_{p}$.
This effect may be important when interpreting results of some accelerator
experiments with high energy unstable particles and also when interpreting
some results of astrophysical observations.
Astrophysical processes are the source of
a huge number of elementary particles including unstable particles
of very high energies.
The numbers of created unstable particles during these processes are
so large that many  of them may survive up to transition times
$t \sim T_{p}$ or much later and they move with ultra relativistic
velocities. From the above discussion it follows that numbers of unstable
particles which survived to these times is much, much greater
than one could expect estimating these numbers by means of the relation (\ref{a-p=a-0}).

The above analysis shows also that the scale and the intensity of
the effect described in \cite{plb-2014} were underestimated there.
In \cite{plb-2014} the instantaneous energy ${\cal E}_{\phi}(t)$ of
an unstable particle $\phi$  was analyzed and it was shown there
that fluctuations of this energy at the transition time region have to occur.
These fluctuations cause changes in the particle velocity which in the case
of charged particles (or particles with the non--zero magnetic moment) forces
them to emit electromagnetic radiation.
The base of estimations performed in \cite{plb-2014} was the
relation of the type (\ref{a-p=a-0}).
Results presented in Fig. \ref{fi2} show that in the case of the moving relativistic particle the
form of these fluctuations seen by the observer ${\cal O}$ is the same as the form o such
fluctuations in the particle rest system but they occur much earlier. What is more
the amplitude of fluctuations of ${\cal E}_{\phi}^{p}(t)$ may be even
larger than the corresponding amplitude of ${\cal E}_{\phi}(t)$ calculated
in the particle rest system. Also the analysis performed in this Section
and results presented in Fig. \ref{fi1} shows that in a real situation much
more unstable particles have to survive up to the transition times
than it can be expected using (\ref{a-p=a-0}) when performing the estimations.
In general one can expect that within the model considered the relation
between true number of the particles, ${\cal N}(t \sim T_{p})  =
{\cal P}_{p}(t \sim T_{p})\,{\cal N}_{0}$, which survived up to
$t \sim T_{p}$, and the corresponding number ${\cal N}(t \sim T_{0}/\gamma_{L})$
obtained assuming (\ref{a-p=a-0}) looks as follows: ${\cal N}(t \sim T_{p})\,
\gtrsim\, 10^{3}\,{\cal N}(t \sim T_{0})$ (compare curves $(a)$ and $(b)$ analyzed
in these Figures for times $t$ belonging to the transition times region and values of
the corresponding survival probabilities). This means that the scale of effect
analyzed in \cite{plb-2014} and its intensity should be  much larger
than it was estimated there.

The last remarks.
There is a remarkable similarity of decay curves presented in Fig.  \ref{fi3} and  results
reported by the  GSI  team   in \cite{litvinov1} and
presented  there in Figs 3 and 5 (for update results see  \cite{kienle1}). The relativistic Lorentz factor in
the GSI experiment was $\gamma_{L} \simeq 1.43$ which is  very
close to the Lorentz factor used in calculations leading to the results presented
in our Fig. \ref{fi1}, Panel $C$,  and Fig. \ref{fi3}.
In Fig. 3 the solid fluctuating decay curve ${\cal P}_{p}(t)$
looks as the curve obtained experimentally by the mentioned GSI team,
whereas the dashed curve being a part of exponentially decreasing
probability ${\cal P}_{0}(t/\gamma_{L})$ at these times $t$ looks
as the expected and calculated theoretically curve  by this team. So one
can not  exclude that choosing an appropriate form of the density
$\omega (m)$ rather different from the simple $\omega_{BW}(m)$ (e.g.  having the form discussed in \cite{giacosa})
and calculating the survival probability ${\cal P}_{p}(t)$
by means of the proper formula (\ref{a-p-spec}) it will be
possible to
reproduce theoretically the
experimental decay curve obtained
by the GSI team and thus to explain the GSI anomaly.

\hfill\\
\noindent
{\bf Acknowledgments:} This work was supported in part
 by the Polish NCN project DEC--2013/09/B/ST2/03455.


\begin{thebibliography}{10}
\bibitem{misra}
B. Misra and E. C. G. Sudarshan, J. Math. Phys. 18,
756 (1977).
\bibitem{chiu1}
C. B. Chiu, B. Misra, and E. C. G. Sudarshan, Phys.
Rev. D 16, 520 (1977);
\bibitem{anti-zeno}
W. C. Schieve, L. P. Horwitz, and J. Levitan, Phys. Lett. A
136, 264 (1989);
A. G. Kofman and G. Kurizki, Nature 405, 546 (2000).
\bibitem{khalfin-2}
L. A. Khalfin, Zh. Eksp. Teor. Fiz. {\bf 33}, 1371 (1957) [Sov.
Phys. --- JETP {\bf 6}, (1958), 1053].
\bibitem{fonda}
L. Fonda, G. C. Ghirardii and A. Rimini,
 Rep. on Prog. in Phys. {\bf 41},  587 (1978).
\bibitem{Wessner}
J. M. Wessner, D. K. Andreson and R. T. Robiscoe, Phys. Rev. Lett.
{\bf 29}, (1972), 1126.
\bibitem{Norman1}
E. B. Norman, S. B. Gazes, S. C. Crane and D. A. Bennet, Phys. Rev.
Lett. {bf 60}, (1988), 2246. E. B. Norman, B. Sur, K. T. Lesko, R.-M. Larimer,
Phys. Lett. {\bf B 357}, (1995), 521.
\bibitem{seke}
J. Seke, W. N. Herfort, Phys. Rev. {\bf A 38}, (1988), 833.
\bibitem{parrot}
R. E. Parrot, J. Lawrence, Europhys. Lett. {\bf 57}, (2002), 632.
\bibitem{lawrence}
J. Lawrence, Journ. Opt. B: Quant. Semiclass. Opt. {\bf 4}, (2002), S446.
\bibitem{joichi}
I. Joichi, Sh. Matsumoto, M. Yoshimura, Phys. Rev. {\bf D 58}, (1998), 045004.
\bibitem{Nowakowski}
N. G. Kelkar, M. Nowakowski and K. P. Khemchandani, Phys. \linebreak Rev.  {\bf C 70}, (2004), 024601.
\bibitem{Nowakowski2}
M. Nowakowski, N. G. Kelkar, arXiv: 0807.5103; AIP Conf. Proc. {\bf 1030}, (2008), 250.
\bibitem{jiitoh}
T. Jiitoh, S. Matsumoto, J. Sato, Y. Sato, K. Takeda, Phys Rev. {\bf A 71},
(2005), 012109.
\bibitem{rothe}
C. Rothe, S. I. Hintschich and A. P. Monkman,
Phys. Rev. Lett. {\bf 96},
163601 (2006).
 \bibitem{plb-2014}
 K. Urbanowski, K. Raczy\'{n}ska, Phys. Letters {\bf B731}, 236 (2014).
\bibitem{exner}
P. Exner, Phys. Rev. D \textbf{28}, 2611 (1983).
\bibitem{khalfin-1}
L. A. Khalfin, {\em Quantum theory of unstable particles and relativity}, PDMI PREPRINT--6/1997 (St. Petersburg Department of Steklov Mathematical Institute, St. Pteresburg, Russia, 1997).
\bibitem{stefanovich}
E. V. Stefanovitch, {\em Quantum effects in relativistic decays}, International Journal of Theoretical Physics, {\bf 35}, 2539 (1996).
\bibitem{shirkov}
M. Shirkov, {\em Decay law of a moving unstable particles}, International Journal of Theoretical Physics, {\bf 43}, 1541 (2004).
\bibitem{Fock}
S. Krylov, V. A. Fock, Zh. Eksp. Teor. Fiz. {\bf 17}, (1947), 93.
\bibitem{nowakowski3}
N. G. Kelkar, M. Nowakowski, J. Phys. A: Math. Theor., \textbf{43}, 385308 (2010).
\bibitem{urbanowski-2008}
K. Urbanowski, Eur. Phys. J. {\bf C 58}, 151, (2008).
\bibitem{gibson}
W. M. Gibson, B. R. Polard, {\em Symmetry principles in elementary particle physics},  Cambridge, 1976.
\bibitem{urbanowski-1-2009}
K. Urbanowski, Eur. Phys. J. \textbf{ D 54},  25 (2009).
\bibitem{pra}
K. Urbanowski, Phys. Rev. A \textbf{50}, 2847 (1994).
\bibitem{litvinov1}
Yu.A. Litvinov {\em et al},
Physics Letters {\bf B 664}, 162 (2008).
\bibitem{kienle1}
P. Kienle {\em et al},
{\em Physics Letters}, {\bf B 726}, 638 (2013).
\bibitem{giacosa}
F. Giacosa, G. Pagliara, Quantum Matter, {\bf 2}, 54, (2013); arXiv: 1110.1669. F. Giacosa, G. Pagliara, {\em (Oscillating) non--exponetial decays of unstable states}, arXiv: 1204.1896.
\end{thebibliography}
\end{document}